\documentclass[pre,twocolumn,letterpaper,showpacs,superscriptaddress,floatfix]{revtex4}
\usepackage{graphicx,latexsym,amsfonts,amsmath,amssymb,bm}

\begin{document}
\title{Diffusion Limited Aggregation with Power-Law Pinning}
\date{\today}
\author{H.G.E Hentschel}
\email{phshgeh@physics.emory.edu}
\affiliation{Department of Physics, Emory University, Atlanta, GA,
30322, USA}
\author{M. N. Popescu}
\email{popescu@mf.mpg.de}
\affiliation{Max-Planck-Institut f\"ur Metallforschung, Heisenbergstr. 3,
D-70569 Stuttgart, Germany}
\affiliation{Institut f\"ur Theoretische und Angewandte Physik,
Universit\"at Stuttgart, Pfaffenwaldring 57, D-70569 Stuttgart, Germany}
\author{F. Family}
\email{phyff@emory.edu}
\affiliation{Department of Physics, Emory University, Atlanta, GA, 30322, USA}

\begin{abstract}
Using stochastic conformal mapping  techniques we study the patterns emerging
from Laplacian growth with a power-law decaying threshold for growth
$R_N^{-\gamma}$ (where $R_N$ is the radius of the $N-$ particle
cluster). For $\gamma > 1$ the growth pattern is in the same universality
class as diffusion limited aggregation (DLA) growth, while for $\gamma < 1$
the resulting patterns have a lower fractal dimension $D(\gamma)$ than a DLA
cluster due to the enhancement of growth at the hot tips of the developing
pattern. Our results indicate that a pinning transition occurs at $\gamma = 1/2$,
significantly smaller than might be expected from the
lower bound $\alpha_{min} \simeq 0.67$ of multifractal spectrum of DLA. This
limiting case shows that the most singular tips in the pruned cluster now
correspond to those expected  for a purely one-dimensional line. Using
multifractal analysis, analytic expressions are established for
$D(\gamma)$ both close to the breakdown of DLA universality class, i.e.,
$\gamma \lesssim 1$, and close to the pinning transition, i.e., $\gamma \gtrsim 1/2$.
\end{abstract}


\maketitle

\section{Introduction}

Nonequilibrium growth models leading naturally to self-organized fractal
structures, such as  diffusion limited aggregation
(DLA) \cite{WS81}, are of continuing interest  due to their relevance for
many important physical processes including
dielectric breakdown~\cite{NPW84}, electrochemical deposition
\cite{BB84,MSHHS84}, and Laplacian flow~\cite{P84}.

A powerful method, namely iterated stochastic conformal mapping \cite{98HL,97Hast},
has been already successfully applied to generate and analyze
DLA \cite{99DHOPSS,BI00} and Laplacian \cite{FB00} growth patterns in two dimensions.
This has provided an alternative way to address many of the important open questions
related to pattern formation in DLA in two dimensions, one of these being the existence
of minimal fields for growth at the boundary of the growing cluster.
In previous work \cite{GH02} we studied the properties induced by a {\it fixed},
material dependent critical field $E_c$ for growth, and showed that in the
presence of such a threshold all clusters ultimately become pinned and, in
addition, this simple constraint has remarkable consequences for the resulting
patterns  -- the rich, branched structure of DLA is replaced by a much
lower dimensional shape consisting of a few surviving branches.

In this paper we  address a similar, but significantly more important
question because of its relationship to the multifractal spectrum of DLA, that
of what happens when there exists a critical field for growth on the boundary
of the cluster, field decaying like $R_N^{-\gamma}$ as the cluster increases
in size. We shall call this model the ``$\gamma$-model''.
As we will show below, as $\gamma$ decreases from $\gamma > 1$ toward a critical
value $\gamma =1/2$ (which corresponds to the most singular possible behavior for
the Laplacian field, that occuring at the tip of a line), there is a continuous
transition from DLA toward lower dimensional shapes for which the multifractal
spectrum is necessarily different from that of DLA.  We study this transition in
terms of the fractal dimension $D(\gamma)$ of the emerging patterns, and we derive
analytic expressions for the  behavior of $D(\gamma)$ in the range
$\gamma \lesssim 1$ where the DLA universality class breaks down and
$\gamma \gtrsim 1/2$ close to the pinning transition.

\section{Model and Theoretical Background}

Witten and Sander \cite{WS81} have shown that the growth probability at
any point $s$ on the boundary of a DLA cluster of length $L$ is given by the
harmonic measure $P(s)~=~|(\nabla V)(s)|/\int_0^L ds' |(\nabla V)(s')|$,
where $V({\bf r})$  obeys Laplace's equation $\nabla^2 V = 0$
subject to the boundary conditions $V = 0$ on the (evolving) boundary of the
cluster and $V \sim \ln r$ as $r \to \infty$ (corresponding to a uniform flux
of particles far away from the cluster).

The model we study is a variant of the two-dimensional DLA growth model
described above in which growth is disallowed at points on the cluster boundary
where the probability for growth is smaller than a critical value $R_N^{-\gamma}$,
where $R_N$ (the exact meaning will be defined later) is the radius of the $N$
particle cluster, i.e.,
\begin{equation}
\label{growth}
P_{grow}(s) =
\begin{cases}
{\displaystyle \frac{|\nabla V (s)|}
{\int_{0}^{L} \theta [\gamma - \alpha (s')]|\nabla V (s')|ds'}},
& |\nabla V| > R_N^{-\gamma},
\cr 0, & |\nabla V| < R_N^{-\gamma},\cr
\end{cases}
\end{equation}
where $\alpha(s')$ is the multifractal exponent at point $s'$ on the cluster boundary,
$L$ is the length of the boundary, and the step function $\theta (\gamma - \alpha (s'))$
ensures that only those regions of the cluster boundary obeying
$|\nabla V(s')|~>~R_N^{-\gamma}$ contribute to the normalization integral.
Estimates of this integral will be very important in our analysis
of the fractal dimension of the growing ``$\gamma$-cluster''. Since we need to
calculate the harmonic measure on a freely evolving interface, this is handled
by using conformal mapping techniques \cite{98HL,99DHOPSS}. The method was
presented in great detail in Refs.~\cite{98HL,99DHOPSS}, and thus here we will
just briefly review the main results.

The basic idea is to follow the evolution of the conformal mapping
$z = \Phi^{(n)}(\omega )$ of the exterior of the unit circle in a mathematical
$\omega$--plane onto the complement of the cluster of $n$ particles in the
physical $z$--plane rather than directly the evolution of the cluster's boundary.
The initial condition is chosen to be $\Phi^{(0)}(\omega ) = \omega$. The process
of adding a new ``particle'' of constant shape and linear scale $\sqrt{\lambda_0}$
to the cluster of $(n-1)$ ``particles'' at a position $s$ which is chosen randomly
according to the harmonic measure is described via a function
$\phi_{\lambda, \theta}(\omega )$, where
\begin{eqnarray}
&&\phi_{\lambda,0}(\omega ) =
\omega^{1-a} \left\{ \frac{(1+ \lambda)}{2\omega}(1+\omega)\right.\nonumber\\
&&\left.\times \left [ 1+\omega+\omega \left( 1+\frac{1}{\omega^2} -
\frac{2}{\omega} \frac{1- \lambda}
{1+ \lambda} \right) ^{1/2} \right] -1 \right \} ^a
\nonumber
\\
&&\phi_{\lambda,\theta} (\omega )
= e^{i \theta} \phi_{\lambda,0}(e^{-i \theta} \omega) \,,
\label{eq-f}
\end{eqnarray}
which conformally maps the unit circle to the unit circle with a bump of linear
size $\sqrt{\lambda}$ localized at the angular position $\theta$ \cite{98HL}.
The shape of the bump depends on the parameter $a$. Following the analysis
in \cite{99DHOPSS}, we have used $a=0.66$ througout this paper, as we believe
the large scale asymptotic properties will not be affected by the microscopic
shape of the added bump.

The conformal map for an $n$-particle cluster $\Phi^{(n)}(\omega )$  can be built
by adding one ``particle'' to an $(n-1)$-particle cluster $\Phi^{(n-1)}(\omega )$,
resulting in the recursive dynamics
\begin{equation}
\Phi^{(n)}(\omega ) = \Phi^{(n-1)}(\phi_{\lambda_n,\theta_{n}}(\omega))
\label{recursive}
\end{equation}
which can be solved in terms of iterations of the elementary
bump map $\phi_{\lambda_{n},\theta_{n}}(\omega )$,
\begin{equation}
\Phi^{(n)}(\omega ) =
\phi_{\lambda_1,\theta_{1}}\circ\phi_{\lambda_2,\theta_{2}}\circ\dots\circ
\phi_{\lambda_n,\theta_{n}}(\omega)\, .
\label{comp}
\end{equation}
In Eqs.~\ref{recursive} and~\ref{comp} the angle $\theta_n \in (0,2\pi]$ at
step $n$ is randomly chosen since the harmonic measure on the real cluster
translates to a uniform measure on the unit circle in the mathematical plane,
\begin{equation}
P(s) ds = \frac{d\theta}{2 \pi},
\label{prob}
\end{equation}
and
\begin{equation}
\lambda_{n} = \frac{\lambda_0}{|{\Phi^{(n-1)}}' (e^{i \theta_n})|^2},
\label{eq-lambda}
\end{equation}
is chosen in order to ensure that the size of the bump in the physical $z$
plane is $\sqrt{\lambda_0}$. Since $\sqrt{\lambda_0}$ is a natural length
scale in the problem, in can be scaled out by measuring all the lengths in
terms of it.
We re-emphasize here that although the composition Eq.~\ref{comp} appears at
first sight to be a standard iteration of stochastic maps, this is not so
because the order of iterations is inverted -- the last point of
the trajectory is the inner argument in this iteration. As a result the
transition from $\Phi^{(n-1)}(\omega )$ to $\Phi^{(n)}(\omega )$ is achieved
by composing the $n$ former maps Eq.~\ref{comp} starting from a different seed.
Finally, identifying~\cite{99DHOPSS} the radius $R_n$ of the growing pattern
with the coefficient $F^{(n)}_1 = \Pi_{i=1}^n (1+ \lambda_i)^a$ in the Laurent
expansion of $\Phi^{(n)}$,
\begin{equation}
\Phi^{(n)}(\omega) = F^{(n)}_{1} \omega + F^{(n)}_{0} + F^{(n)}_{-1}\omega^{-1} +
F^{(n)}_{-2}\omega^{-2} + \dots ~,
\label{eq-laurent-F}
\end{equation}
the constraint to grow only at values of $\theta$ which obey Eq.~\ref{growth}
translates into
\begin{equation}
\label{constraint}
\frac{1}{|[\Phi^{(n-1)}]'(e^{i\theta})|} > \left({F_1^{(n-1)}}\right)^{-\gamma}~.
\end{equation}
This constraint is implemented as follows. At step $n$, $\theta_n$ is choosen
from a uniform distribution in $(0, 2\pi]$, independent of previous history.
If it obeys the constraint given by Eq.~\ref{constraint} accept this value
of $\theta_n$, otherwise repeat until the constraint is obeyed.

\section{Results and Discussion}

For the model defined in Sec.~II one would expect the resultant patterns to have
fractal shapes which depend on $\gamma$, and in order to characterize these shapes
we will focus on the scaling behavior of the first Laurent coefficient $F_1^{(n)}$.
Following the arguments in Ref.~\cite{99DHOPSS}, for a given value $\gamma$ one
expects a scaling law of the form
\begin{equation}
\label{F1_scaling}
F_1^{(n)} \sim n^{1/D(\gamma)}~,
\end{equation}
where $D(\gamma)$ is the effective fractal dimension of the resulting cluster.

\begin{figure}[!thb]
\begin{minipage}[c]{.95\columnwidth}
\includegraphics[width=.85\textwidth]{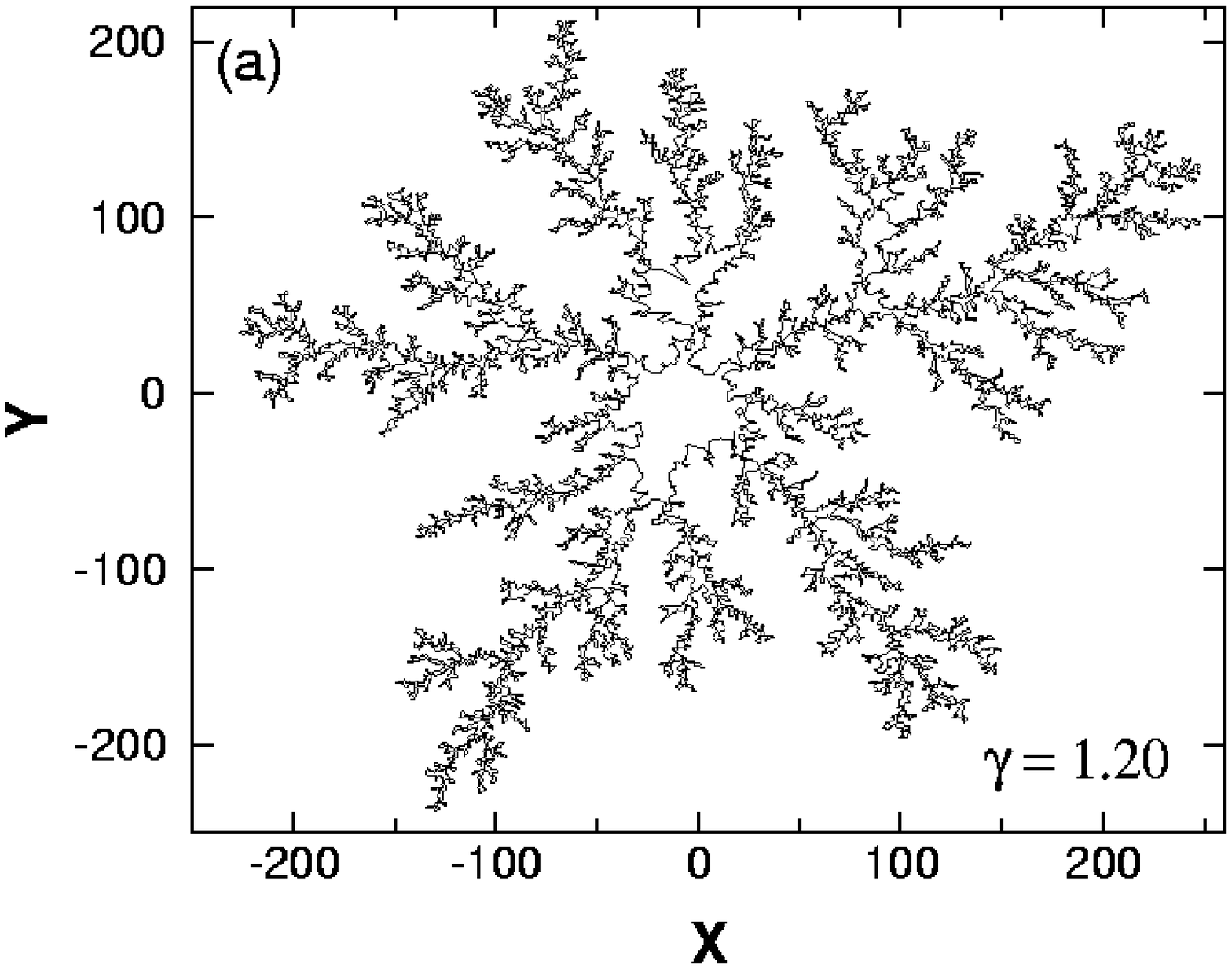}
\end{minipage}%

\begin{minipage}[c]{.95\columnwidth}
\includegraphics[width=.85\textwidth]{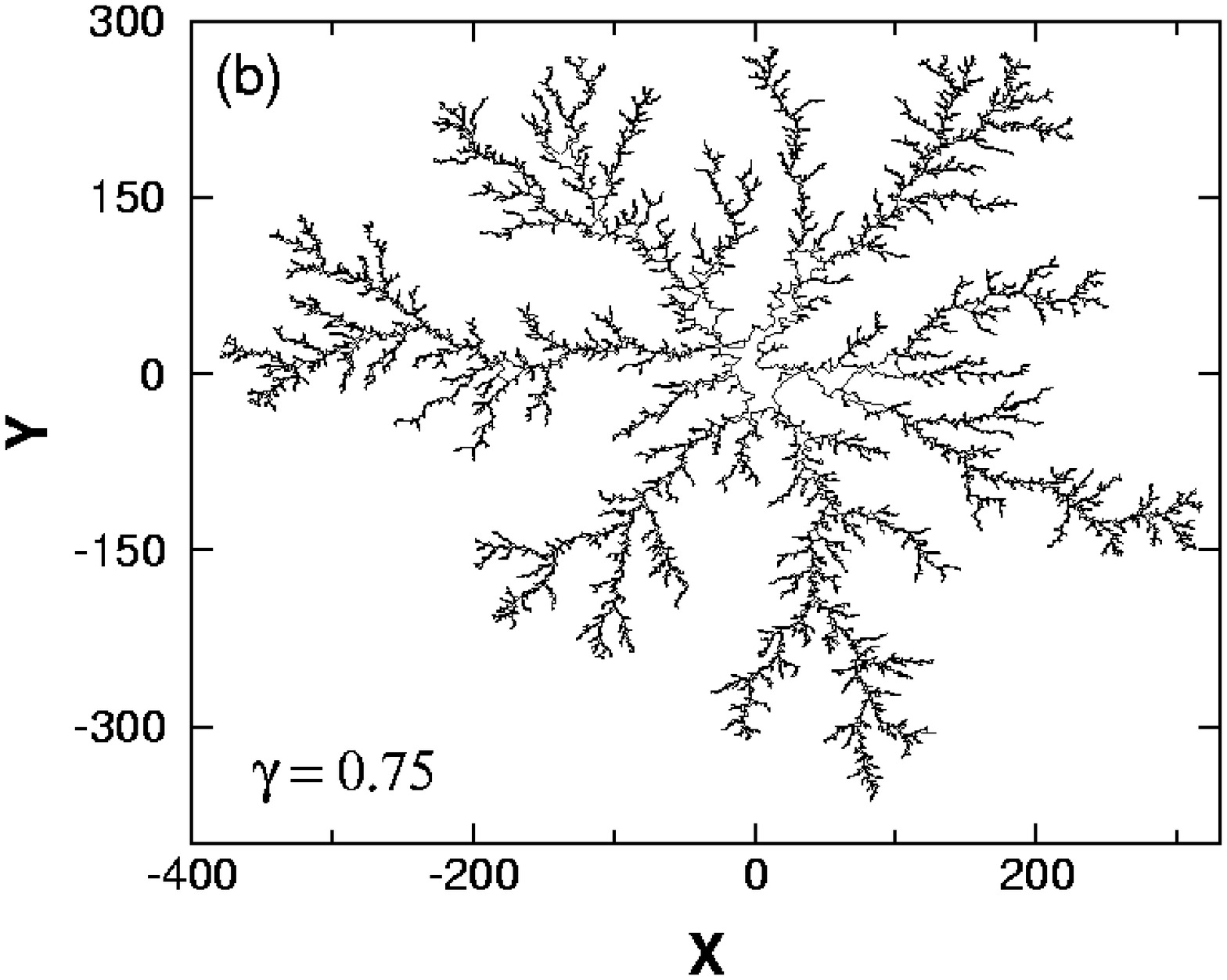}
\end{minipage}%

\begin{minipage}[c]{.95\columnwidth}
\hspace{.1in}\includegraphics[width=.9\textwidth]{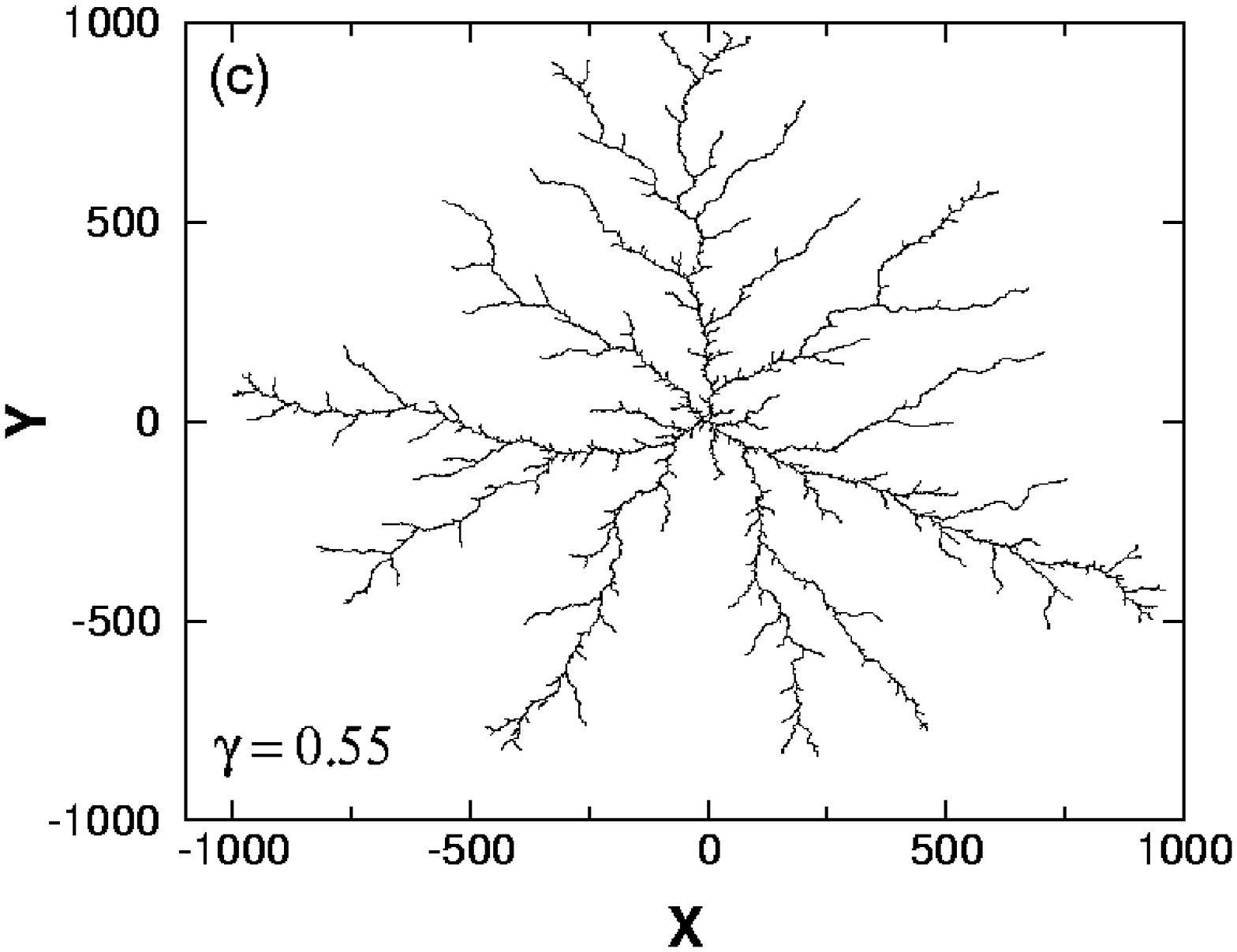}
\end{minipage}

\caption{
\label{fig1}
Typical clusters (size $N=40000$) grown with (a)~$\gamma = 1.20$,
(b)~$\gamma = 0.75$, and (c)~$\gamma = 0.55$, respectively.}
\end{figure}

We have simulated the model defined in Sec.~II for a number of values $\gamma$
in the range $1/2 \leq \gamma \leq 1.2$ and we have calculated $F_1^{(n)}$ as an
average over $100$ clusters (for $\gamma > 0.65$), and, respectively, over $20$
clusters (for $\gamma \leq 0.65$) of size $N = 40000$. In Fig.~\ref{fig1} we show
typical clusters of size $N = 40000$ for $\gamma = 1.20$, $0.75$ and $0.55$,
respectively.
It can be easily seen that the rich, branched structure of the cluster
at $\gamma = 1.20$ changes, as $\gamma$ decreases toward $\gamma = 0.55$,
into a much lower-dimensional shape with only a few branches surviving.
This can be intuitively understood by considering the effect the pinning
probability $R_n^{-\gamma}$ has on the multifractal spectrum of the cluster.
From multifractal scaling~\cite{83HP,86HMP,86HJKPS} we know that the interface
of a fully developed DLA cluster consists of sets of
${\cal N}_{DLA}(\alpha)\sim R_n^{f_{DLA}(\alpha )}$ sites with growing probabilities
$|{\bf \nabla }V|\sim R_n^{-\alpha}$, and we will assume that such a structure
is also valid for the clusters grown with pinning probability $R_n^{-\gamma}$.
As growth proceeds, the lowest probability sites (large $\alpha$), normally
deep in the fjords, will be pinned first, with the hot tips surviving longest,
leading to ``pruning'' of the branches where the tips have a singularity
$\alpha > \gamma$.

\begin{figure}[!thb]

\begin{minipage}[c]{.95\columnwidth}
\includegraphics[width=.82\textwidth]{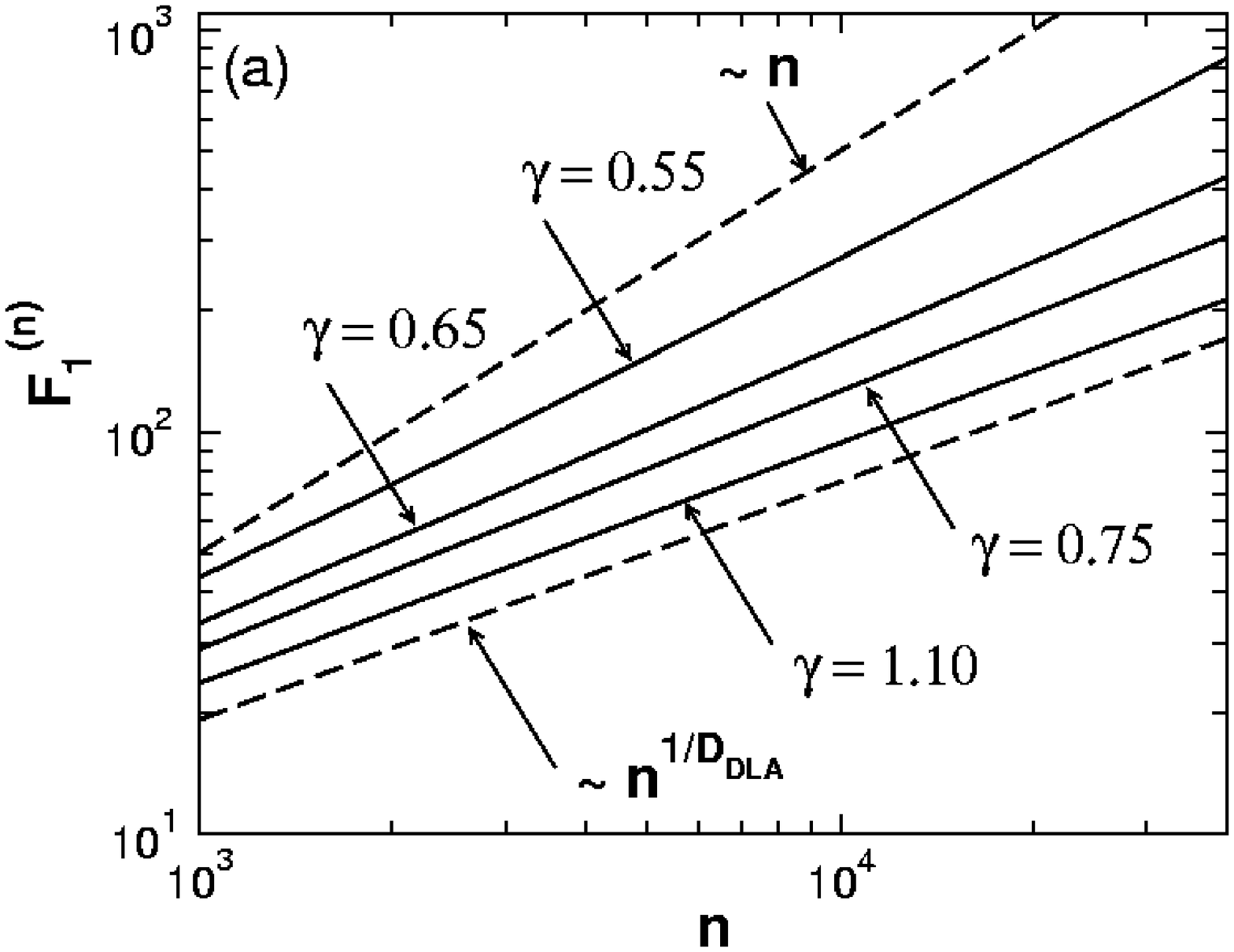}
\end{minipage}%

\begin{minipage}[c]{.95\columnwidth}
\includegraphics[width=.85\textwidth]{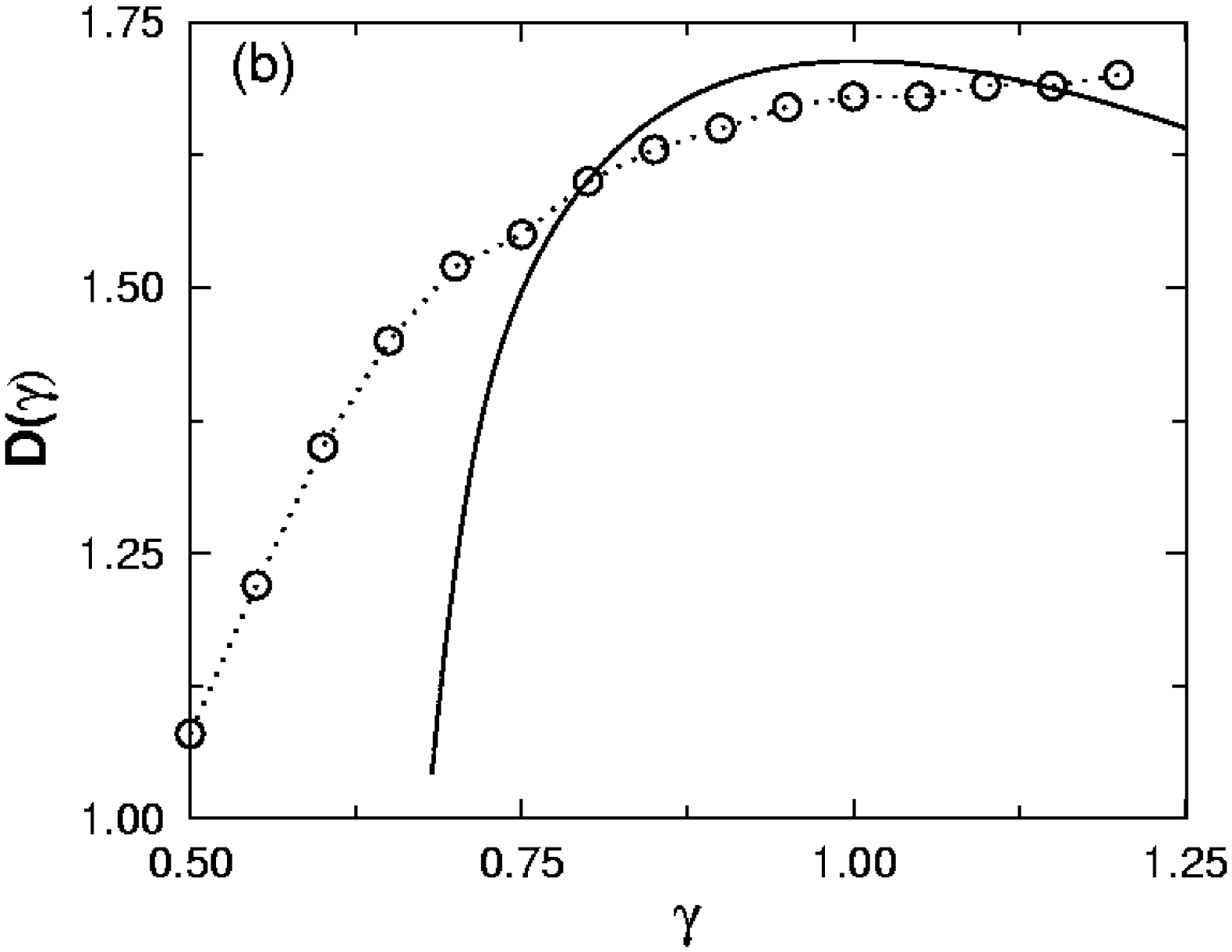}
\end{minipage}%

\caption{
\label{fig2}
(a) $F_1^{(n)}$ as a function of $n$ for clusters grown with
$\gamma = 1.20, 0.75, 0.65, \textrm{and } 0.55$, respectively (log-log plot).
Also shown (dashed lines) are the limit cases of a DLA cluster,
$F_1^{(n)} \sim n^{1/D_{DLA}}$, where $D_{DLA} = 1.71$, and of a line
cluster, $F_1^{(n)} \sim~ n$, respectively. (b) The effective fractal
dimension $D(\gamma)$ obtained from $F_1^{(n)} \sim n^{1/D(\gamma)}$
as a function of $\gamma$ (circles). The dotted line is just a guide
to the eye. The solid line is the theoretical prediction (Eq.~\ref{D_gam})
with $f_\gamma(\gamma) = f_{DLA}(\gamma)$ from Ref.~\cite{MJ02}.
}
\end{figure}

As anticipated, for all the values $0.5 \leq \gamma \leq 1.2$ that we have
tested the coefficient $F_1^{(n)}$ has a clear power-law dependence on the
size $n$, as shown in Fig.~\ref{fig2}(a). Assuming the exponent $1/D(\gamma)$
to be related to a fractal dimension as given by Eq.~\ref{F1_scaling}, the
dependency $D(\gamma)$ (shown in Fig.~\ref{fig2}(b)) is obtained from a
power-law fit to the data.
It can be seen that at  values of $\gamma \gtrsim 1$ the behavior is close to
that of a DLA cluster, i.e., $D(\gamma > 1) \to D_{DLA} \simeq 1.71$, while
for $\gamma \geq 1/2$ the behavior of the radius $F_1^{(n)}$ tends to $n$,
i.e., $D(\gamma \to 1/2) \to 1$, thus the behavior of a growing line.

In order to understand these results theoretically let us begin with a
very simple argument based on the assumption that the clusters have a
multifractal spectrum in the sense described above, in that the interface
consists of sets of ${\cal N}_\gamma(\alpha)\sim R_n^{f_{\gamma}(\alpha )}$
sites with growth probabilities $|\nabla V|\sim R_n^{-\alpha}$.
Since the constraint will cut-off growth at regions in the cluster with
exponents in the range $\alpha > \gamma $ of the multifractal spectrum,
we can write down the following equation for the rate of growth of the
cluster in the presence of the barrier in terms of the rate of growth
of a DLA (no barrier for growth) cluster
\begin{equation}
\label{dRdN}
\left(\frac{dR}{dN}\right)_\gamma \sim \left(\frac{dR}{dN}\right)_{DLA}
\times
\left(\int_{\alpha_{min}}^{\,\gamma} \,d\alpha\, C(\alpha)\,
R^{f_{\gamma}(\alpha )-\alpha}\right)^{-1},
\nonumber
\end{equation}
where $f(\alpha_{min})=0$. The enhancement of growth comes from Eq.~\ref{growth}
together with the estimate
\begin{equation}
\label{estimate}
\int_{0}^{L}|\theta (\gamma - \alpha (s))|\nabla V |(s')ds'\sim
\int_{\alpha_{min}}^{\,\gamma} \,d\alpha\, C(\alpha)\,
R^{f_{\gamma}(\alpha )-\alpha}.
\end{equation}
Now, we know that the harmonic measure is
concentrated at $\alpha = 1$, and thus for $\gamma > 1$ the integral
is dominated by the value of the integrand at $\alpha = 1$, while for
$\gamma \lesssim 1$ it is dominated by the value at $\gamma$, and thus
\begin{equation}
\label{near_one}
\left(\frac{dR}{dN}\right)_\gamma \sim \left(\frac{dR}{dN}\right)_{DLA}
\times
\begin{cases}
K(\gamma ),  & ~\textrm{for}~\gamma > 1, \cr
R^{\gamma-f_{\gamma}(\gamma)}, & ~\textrm{for}~\gamma \lesssim 1,\cr
\end{cases}
\end{equation}
where $K(\gamma )$ is some constant independent of $N$. Eq.~\ref{near_one}
therefore implies
\begin{equation}
\label{D_gam}
D(\gamma) =
\begin{cases}
D_{DLA},  & ~\textrm{for}~\gamma > 1, \cr
D_{DLA} + f_{\gamma}(\gamma ) - \gamma, & ~\textrm{for}~\gamma \lesssim 1.\cr
\end{cases}
\end{equation}

For the case where $\gamma \lesssim 1$, i.e., close to the breakdown in the
DLA universality class, we may assume that the multifractal spectrum of the
cluster is only weakly perturbed from its value in DLA and therefore
$f_{\gamma}(\gamma )\approx f_{DLA}(\gamma )$, or
\begin{equation}
\label{D_gam_2}
D(\gamma) =
\begin{cases}
D_{DLA},  & ~\textrm{for}~\gamma > 1, \cr
D_{DLA} + f_{DLA}(\gamma ) - \gamma, & ~\textrm{for}~\gamma \lesssim 1.\cr
\end{cases}
\end{equation}

This prediction may be tested using the recently computed
$f_{DLA}(\alpha)$ spectrum \cite{MJ02}. The results are shown in
Fig.~\ref{fig2}(b) (solid line), and it can be seen that for
$\gamma \simeq 1$ the theoretical predictions are indeed close to the
measured values $D(\gamma)$. The discrepancies at smaller values of $\gamma$
can be attributed to the fact that the multifractal spectrum of the cluster
is not exactly the one of DLA, and when $\gamma \rightarrow 1/2$, as we will
now discuss, it actually may be expected to deviate significantly from that
of a DLA.

\begin{figure}[!thb]

\begin{minipage}[c]{.95\columnwidth}
\includegraphics[width=.85\textwidth]{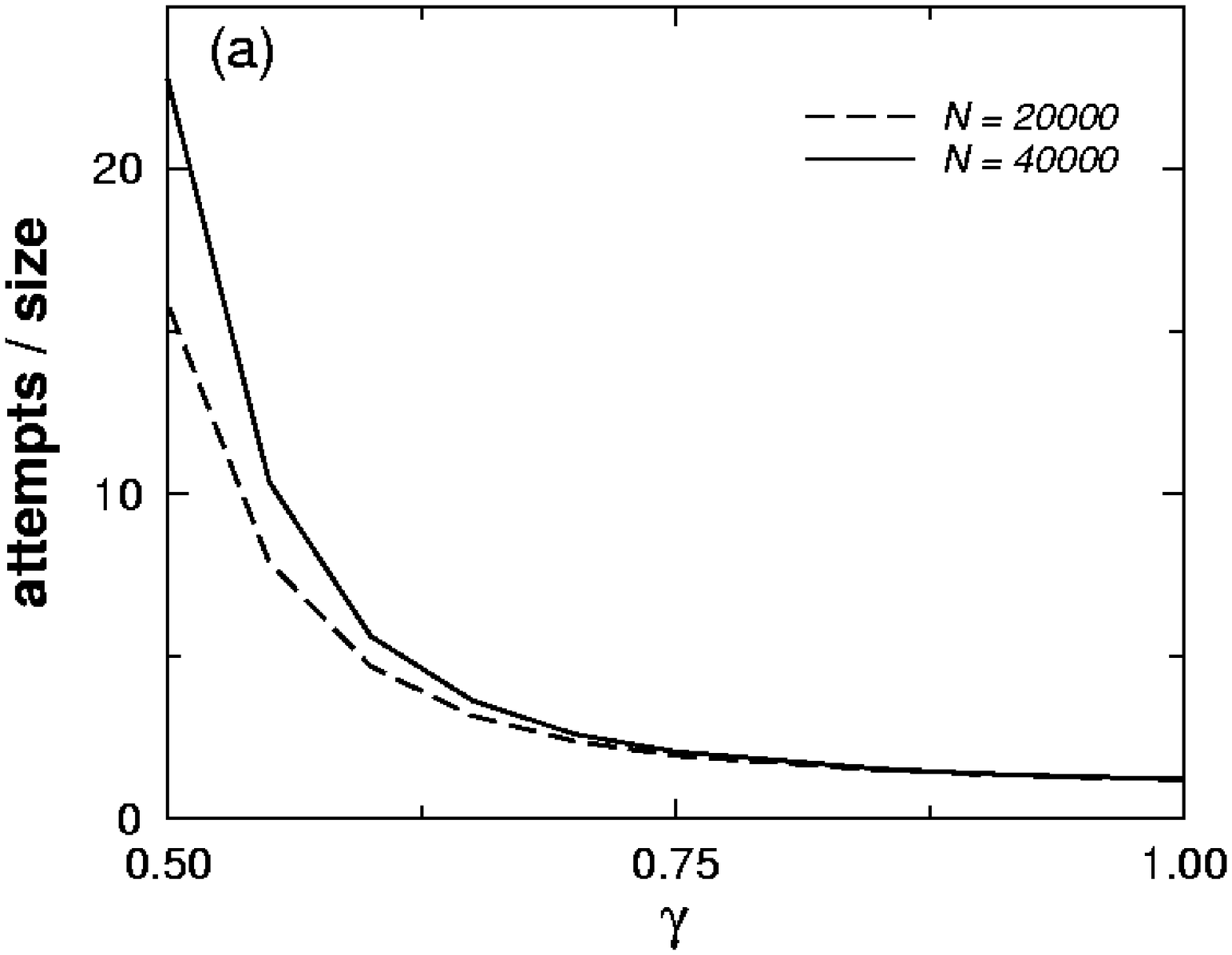}
\end{minipage}%

\begin{minipage}[c]{.95\columnwidth}
\includegraphics[width=.85\textwidth]{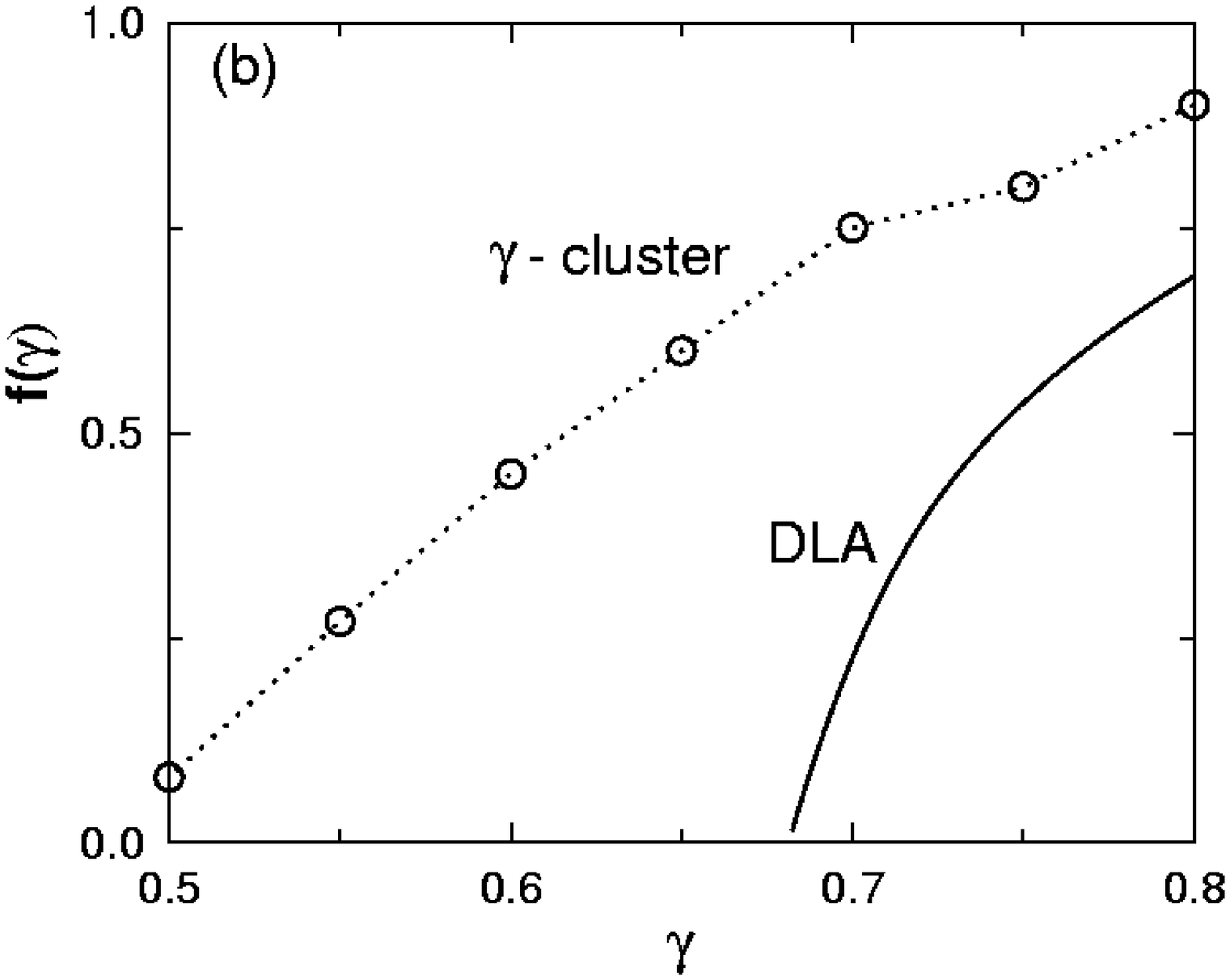}
\end{minipage}%

\caption{
\label{fig3}
(a) Average number of attempts for growing clusters of sizes $N=20000$
(dashed line) and $N=40000$ (solid line), scaled by the actual size of
the cluster, respectively, as function of $\gamma$ (b) The multifractal
spectrum $\tilde f(\gamma)$ from Eq.~\ref{tildef} (symbols) in comparison
with the multifractal spectrum of DLA (solid line) from Ref.~\cite{MJ02}.}
\end{figure}

For $\gamma \gtrsim 1/2$, the change in the multifractal spectrum from that
for DLA is significant. For example, it is known \cite{85TS,MJ02} that for the
DLA spectrum $\alpha_{min} \simeq 0.67$, while we see that  growth continues
significantly below this value, with $\alpha_{min} = 1/2$ being the asymptotic
limit. Our simulations show strong evidence for this limit, as can be seen in
Fig.~\ref{fig3}(a): the average number of attempts for growing clusters of
sizes $N=20000$ and $N=40000$ (scaled by the actual size of
the cluster) exibits a steep increase as $\gamma \to 0.5$.

We shall assume that  for these highly pruned ``$\gamma$-clusters'' there
is a well defined limiting form for the multifractal spectrum,
$f_{\gamma}(\alpha )$, defined for $1/2<\alpha <\gamma$ and obeying
$f_\gamma(1/2)=0$, which is a monotonically increasing function of $\gamma$.
Defining $\tilde f(\gamma) = f_{\gamma}(\gamma )$, we can use a reasoning similar
to that used by Turkevich and Sher~\cite{85TS} in their estimate of the fractal
dimension of DLA to estimate $\tilde f(\gamma)$. The idea is that for
$\gamma \gtrsim 1/2$ only the ``hottest'' tips contribute to growth, and thus one
can write
\begin{eqnarray}
\label{near_0}
\left(\frac{dR}{dN}\right)_\gamma & \sim & P_{max} \sim R^{-1/2}
\times \left(\int_{1/2}^{\,\gamma} \,d\alpha\, C(\alpha)\,
R^{\tilde f(\alpha)-\alpha}\right)^{-1}
\nonumber
\\
&\sim & R^{-1/2-\tilde f(\gamma)+\gamma},
\end{eqnarray}
where the last relation follows from the fact that the integral is dominated
by the value of the integrand at $\gamma$. Thus, we obtain
\begin{equation}
\label{tildef}
\tilde f(\gamma) = D(\gamma) + \gamma - 3/2.
\end{equation}
Since from simulations we know the values $D(\gamma)$, Eq.~\ref{tildef}
(which is valid for $\gamma \gtrsim 0.5$) allows the calculation of the upper
multifractal exponent $\tilde f(\gamma)$. As shown in Fig.~\ref{fig3}(b),
the pinning threshold $R^{-\gamma}$ leads to a shifting of the multifractal
spectrum for $\alpha < \gamma$ to the left, i.e.,
$f_{\gamma}(\alpha )> f_{DLA}(\alpha )$ (more hot tips, and larger fields at
those hot tips, due to pruning).

These results can be intuitively understood as a flow of singularities away
from $\gamma$ (which acts as an unstable fixed point of the dynamics). For any
particular value $\alpha_0 < \gamma$, what happens  while the cluster evolves is
that screening is reduced compared to DLA and therefore there is a flow of
singularities $\alpha_0 \rightarrow \alpha_1$ with $\alpha_1 < \alpha_0$.
In addition, new singularities with $\alpha < \alpha^{(DLA)}_{min}$, can be
created. Thus,  we would expect that the number of singularities
${\cal N}_{\gamma}(\alpha_1)\approx {\cal N}_{DLA}(\alpha_0)$ or
\begin{equation}
\label{flow}
f_{\gamma}(\alpha_1(\alpha_0 )) \approx f_{DLA}(\alpha_0).
\end{equation}
On the other hand for $\alpha_0 > \gamma$ the singularity flow
$\alpha_0 \rightarrow \alpha_1$ can only act toward an increase
$\alpha_1 \geq \alpha_0$
since such points can never grow and thus can only either keep their original
singularity or get a higher value of $\alpha$ during growth.

\section{Conclusions}

Using the stochastic conformal mapping  techniques we have studied the
patterns emerging from Laplacian growth with a power-law decaying threshold
for growth $R_N^{-\gamma}$, $\gamma \geq 1/2$. We have shown that due to the
enhancement of growth at the hot tips as $\gamma$ decreases the growth evolves
from  patterns in the DLA universality class for $\gamma > 1$ to clusters with
a lower fractal dimension $D(\gamma)$ for $\gamma < 1$ due to the enhancement
of growth at the hot tips.
We have presented evidence that $\gamma = 1/2$, corresponding to the
singularities at the tip of a purely one-dimensional (line) growth pattern,
is the lower limit for growth, with all clusters becoming ultimately pinned for
$\gamma < 1/2$. By using multifractal analysis, we have proposed analytic
expressions for $D(\gamma)$ for both $\gamma \lesssim 1$ near the breakdown of
the DLA universality class and near the pinning transition $\gamma \gtrsim 1/2$.
Finally, we have shown that in the small $\gamma$ range the multifractal spectrum
of the resulting cluster is significantly changed from that of a DLA. We have
suggested that this change may be due to a flow of singularities with $\gamma$
acting as an unstable fixed point of the dynamics,
but further work will be necessary to fully elucidate this point.

\begin{acknowledgments}

This work has been supported by the Petroleum Research Fund.
We would like to acknowledge the very stimulating discussions, leading
to the idea for  this paper with Tom Witten, Benny Davidovitch and Thomas
Seligman and  at the
CIC, UNAM in Cuernavaca, Morelos, Mexico.
One of us (MNP) would also like to thank the Physics Department at
Emory University for hospitality during the period when some of this
work was done. The authors would like to thank Prof. I. Procaccia and
the authors of Ref.~\cite{MJ02} for making available the data for the
multifractal spectrum of DLA.

\end{acknowledgments}


\end{document}